\documentclass[preprint,12pt]{elsarticle}
\usepackage{graphicx} 
\usepackage{epsfig}
\usepackage{amssymb}
\usepackage{amsmath}
\usepackage{bm}
\biboptions{square,compress}

\journal{Physica A}

\begin{document}

\begin{frontmatter}

\title{Steady-state distributions of probability fluxes on complex networks}

\author{Przemys{\l}aw Che{\l}miniak\corref{cor1}}
\ead{geronimo@amu.edu.pl}
\author{Micha{\l} Kurzy\'nski}
%\ead{kurzphys@amu.edu.pl}

\cortext[cor1]{Corresponding author}

\address{Faculty of Physics, A. Mickiewicz University, Umultowska 85, Pozna\'n, Poland}

\begin{abstract}
The methodology based on the random walk processes is adapted and applied to a comprehensive
analysis of the statistical properties of the probability fluxes. To this aim we define a simple
model of the Markovian stochastic dynamics on a complex network extended by the additional
transition, called hereafter the gate. The random skips through the gate, driven by the external
constant force, violate the detailed balance in the network. We argue, using a theoretical
approach and numerical simulations, that the stationary distributions of the probability
fluxes emergent under such conditions converge, regardless of the network topology, to the
normal distribution. This result, combined with the stationary fluctuation theorem, permits to
show that its standard deviation depends directly on the square root of the average flux. In
turn, the central result of our paper relates this quantity to the external constant force
and the two parameters that entirely characterize the normal distribution of the probability
fluxes both close to as well as far from the equilibrium state. Also, the other effects that
modify these parameters, such as the addition of shortcuts to the tree-like network, the
extension and configuration of the gate and a change in the network size studied by means of
the computer simulations are widely discussed in terms of the rigorous theoretical predictions.
\end{abstract}


\end{frontmatter}

%\linenumbers

\section{Introduction}
\label{sec1}

\begin{figure}[t]
\begin{center}
\includegraphics[scale=0.35]{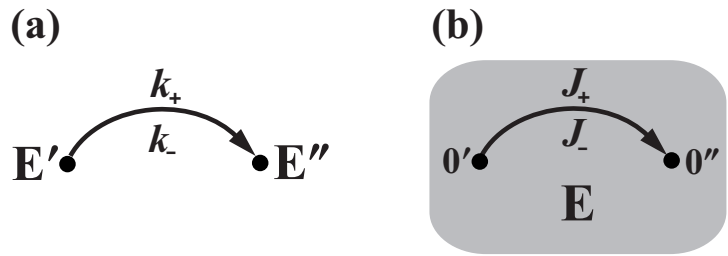}
\end{center}
\vspace{-0.5cm}
\caption{(a) The scheme of the enzymatic reaction that proceeds at the rate constants $k_{+}$ and
$k_{-}$ between two distinguished conformational states $E'$ and $E''$ of the protein enzyme.
(b) The reaction fluxes $J_{+}$ and $J_{-}$ through the gate that consists of two conformational
substates $0'$ and $0''$. The gray plaque represents the network of conformational transitions
within either the enzyme or the enzyme-substrate native state $E$. Arrows indicate the directions
assumed to be forward.}
\label{fig_1}
\end{figure}

According to the principles established by the conventional enzymology, a protein that
catalyzes a chemical reaction occurs in a few most stable conformational states~\cite{
Fersht1999}. The most simplified version of this paradigm is sketched in Fig.~\ref{fig_1}a,
where the enzyme occurs in only two distinguished states $\mathrm{E}'$ and $\mathrm{E}''$.
The transitions between them proceed at the rate constants $k_{+}$ and $k_{-}$, depending
on the molar concentrations of substrates taking part in this process. In a non-equilibrium
steady state, the average reaction flux~\cite{Kurzynski2003,Kurzynski2006}
\begin{equation}
\label{eq_1}
J=\frac{1-e^{-a}}{k_{+}^{-1}+k_{-}^{-1}e^{-a}}
\end{equation}
is unbalanced, because of the dimensionless constant force $a\!\equiv A/k_{\mathrm{B}}T$,
defined by the ratio of the thermodynamic force (affinity) $A$, and the temperature
$T$ multiplied by the Boltzmann constant $k_{\mathrm{B}}$. In Fig.~\ref{fig_1}b, this
oversimplified picture of the coarse-grained enzymatic kinetics is replaced by the more
detailed 'mesoscopic' scheme, promoted in our previous papers~\cite{Kurzynski2014a,
Kurzynski2014b}. The gray rectangle represents an arbitrary network of stochastic transitions
between numerous conformational substates, composing either the enzyme or the enzyme-substrate
native state E. All these internal transitions satisfy the detailed balance condition.
However, it is broken by the external transitions powered by the chemical reaction that
follows with the rate constants (reaction fluxes) $J_{+}$ and $J_{-}$ through a single
pair (the 'gate') of conformational states $0'$ and $0''$. Unlike the average flux in
Eq.~\ref{eq_1}, the fluxes emergent in a such mesoscopic system fluctuate in time~\cite{
Sekimoto2010,Seifert2008,Seifert2010}. Thus, they must be considered as the random
variables~\cite{Gardiner2004,Mahnke2009}, and their statistical properties provide the scope
of the present paper.

For this purpose, we construct a dynamical system like this, drafted in Fig.~\ref{fig_1}b, in
which the stationary fluxes are driven by a time-independent external force. In our model the
states of the system, symbolized by the nodes, are assumed to be organized into a complex network,
while the stochastic transitions proceed between them through the links~\cite{Newman2010,Barrat2008}.
The additional transitions that violate the detailed balance in the network take place through the
gate which consists of two arbitrarily chosen nodes, not necessarily connected by a direct link. All
these transitions are determined by a set of master equations~\cite{Gardiner2004}. A lot is known
about its analytical solutions on the simple networks, e.g. the regular or fractal lattices, having
defined the Euclidean or the fractal dimension, respectively. Nevertheless, the model we propose in
this paper is more general and can be applied to any network with Markovian stochastic dynamics.

The central result of our paper is the relation like that in Eq.~\ref{eq_1}. We show that the
two parameters present in this formula are completely sufficient to specify the basic properties
of the probability fluxes. In the non-equilibrium ensemble these quantities are no longer
determined precisely, but undergo the stationary distributions constrained by the fluctuation
theorem in the Andrieux-Gaspard form~\cite{Andrieux2006,Andrieux2007c}. In general, it relates
the probability to observe variations of a certain fluctuating variable proceding forward in
time to that of observing its change in a time-reversed process~\cite{Evans2002,Jarzynski1997,
Kurchan1998, Lebowitz1999,Crooks1999,Hummer2001,Hatano2001,Andrieux2007,Sagawa2012} and permits
to comprehend how irreversibility emerges from the reversible dynamics~\cite{Feng2008,Jarzynski2008,
Jarzynski2011}. The correctness of the different variants of the fluctuation theorem have been
verified in the case of many experimental protocols~\cite{Ritort2007,Toyabe2010,Trepagnier2004,
Liphardt2002,Collin2005,Bustamante2005}, as well as theoretical models~\cite{Bena2005,Cleuren2006a,
Cleuren2006b,Crooks2007,Lu2014,Dahr2005,Ritort2002,Speck2005,Jarzynski1999,Seifert2011,Mandal2013,
Evans1993,Searles1999,Dellago2014}, for the systems that operate on the mesoscopic scales. Here,
we use the method of numerical simulations and argue that the stationary distributions of the
probability fluxes converge to the Gaussian function. Its width, or equivalently the standard
deviation, depends directly on the mean flux for the given values of the external force and the
time period of the fluxes determination. Also, the other effects that influence the statistical
properties of the probability fluxes, such as the addition of shortcuts to the tree-like network,
the extension of the gate, its configuration on the network and the network size, are thoroughly
studied and interpreted in terms of these rigorous theoretical predictions. We summarize our
results in the final part of this paper.

\section{Stochastic dynamics on networks. The case of a single gate}
\label{sec2}

A continuous-time evolution of the Markovian stochastic process in a discrete set
of states is described by the system of differential master equations~\cite{Gardiner2004,
Mahnke2009}
\begin{equation}
\label{eq_2}
\dot{p}_{l}(t)\!=\!\sum_{l'}\bigl[w_{ll'}p_{l'}(t)-w_{l'l}p_{l}(t)\bigr],
\end{equation}
where $p_{l}(t)$ refers to the occupation probability of a state $l$ at time
$t$, the over-dot denotes a derivative with respect to time $t$ and $w_{l'l}$ are
the transition probabilities per unit time from the state $l$ to the adjacent
states $l'$. In general, these transition rates need not complement to unity after
summation over $l'$ for a fixed $l$, which poses some technical problems for numerical
simulations. However, on introducing a discrete time $t\!=\!m\tau_{0}$, where the
number of steps $m$ is expressed in the specified unit of time $\tau_{0}$, Eq.~\ref{eq_2}
can be rewritten as the difference master equation
\begin{equation}
\label{eq_3}
p_{l}(m+1)\!=\!p_{l}(m)+\sum_{l'}\bigl[u_{ll'}p_{l'}(m)-u_{l'l}p_{l}(m)\bigr],
\end{equation}
with the transition probabilities
\begin{equation}
\label{eq_4}
u_{l'l}\!=\!\tau_{0}w_{l'l}
\end{equation}
being now properly normalized to unity
\begin{equation}
\label{eq_5}
\sum_{l'}u_{l'l}=1.
\end{equation}
In general, the above summation also includes the probability of waiting $u_{ll}$ on a given
node $l$, that determines in fact $\tau_{0}$. They cancel out in Eqs.~\ref{eq_2} and \ref{eq_3},
thanks to which the dynamics remain invariant due to such modifications. In its dimensionless
form, Eq.~\ref{eq_3} is ready-to-use for the numerical applications we deal with in the separate
section.

For the purpose of our further considerations, we already assume for now that the set of states,
symbolized by the nodes (vertices), is to be organized into an arbitrarily complex network.
From this point of view Eq.~\ref{eq_2} is equivalent to a graph in which the random transitions
between the nodes proceed through the system of links (edges)~\cite{Newman2010,Barrat2008}. We
point out that a specific trait of our model is a pair of distinguished nodes $l\!=\!0'$, $0''$
that compose the gate through which additional transitions can occur. On the whole, these two
appointed nodes need not be connected by a direct link. All that we have supposed so far on our
model leads to the following formula for the transition rates in Eq.~\ref{eq_2}:
\begin{equation}
\label{eq_6}
w_{l'l}=v_{l'l}+v_{+}\delta_{l'0''}\delta_{l0'}+v_{-}\delta_{l'0'}\delta_{l0''},
\end{equation}
where $\delta_{ll'}\!=\!1$ for $l\!=\!l'$ and 0 otherwise. The quantities $v_{l'l}$ are
the internal transition probabilities per unit time from the node $l$ to its nearest
neighbours $l'$, which in equilibrium (the superscript "eq") satisfy the detailed balance
condition
\begin{equation}
\label{eq_7}
v_{ll'}p_{l'}^{\mathrm{eq}}=v_{l'l}p_{l}^{\mathrm{eq}}.
\end{equation}
By analogy with the transition state theory ~\cite{Kurzynski2006,Truhlar1996}, both
the left and the right side of the above equation correspond to the reciprocal mean
transition time
\begin{equation}
\label{eq_8}
\frac{1}{\tau_{l'l}}=v_{l'l}p_{l}^{\mathrm{eq}}
\end{equation}
from the node $l$ occupied with the probability $p_{l}^{\mathrm{eq}}$ to the adjacent node
$l'$ (or {\it vice versa}). Accordingly, we can define the external transition rate through
the gate from the node $0'$ to $0''$
\begin{equation}
\label{eq_9}
v_{+}=\frac{1}{\tau p_{0'}^{\mathrm{eq}}}
\end{equation}
that within our model is assumed to proceed in the forward direction, whereas
\begin{equation}
\label{eq_10}
v_{-}=\frac{e^{-a}}{\tau p_{0''}^{\mathrm{eq}}}
\end{equation}
corresponds to the external passage through the same gate in a backward direction,
namely from the node $0''$ to $0'$. The external mean transition time $\tau$ is to be
compared with the internal mean transition times $\tau_{l'l}$. The time $\tau$ and the
dimensionless thermodynamic force $a$ (see Eq.~\ref{eq_1}) are the only parameters that
determine the rates of external transitions through the gate, whereas the exponential
factor $e^{-a}$, appearing in Eq.~\ref{eq_10}, breaks the detailed balance symmetry in
Eq.~\ref{eq_7}.

When a dynamical system is kept away from the equilibrium by the action of a time-independent
external force, then, after a certain transient period of time, it attains the stationary
state. We do not intend to debate here whether this state is stable or unstable. Important,
however, is that under such conditions, the stochastic transitions between its microscopic
states begin to violate the detailed balance condition. Consequently, the non-vanishing
stationary currents or fluxes will arise in a whole system. The same scenario concerns
the complex network with the stochastic dynamics defined in the next section. As the
dynamical system, its state is identified with a node $l$ occupied at a given instant
of time with the probability $p_{l}(t)$ that evolves in time, according to the law of motion,
determined by the set of the coupled master equations (see Eq.\ref{eq_2}). In the following,
we want to predict how these dynamics stipulate the steady-state distributions of the
probability fluxes through the external gate. 

\section{Fractal scale-free networks with specified dynamics of transitions between nodes}
\label{sec3}

The structural properties of any complex network are characterized by a few essential
parameters~\cite{Newman2010}. The total number of nodes $N$ determines the size of
a network, whereas a density of internode connections depends on the total number
of links $L$. An alternative measure of the network size is the mean shortest distance
between any two nodes, defined as the length of the shortest path through the network links
averaged over all the pairs of nodes. The basic property of each node labeled by $l$ is
its degree $k_{l}$, i.e. the number of links it has to the neighbouring nodes. The emergent
feature of the entire network is the degree distribution function $P(k)$ which provides
the probability that the randomly selected node has exactly $k$ edges. A determination
of this function allows us to calculate the average degree and recognize the many other intrinsic
properties of the network structure~\cite{Barabasi1999}. For the regular networks, the degree
distribution function displays a single sharp spike, because the majority of nodes, excepting
the boundary ones, have the same number of the nearest neighbours. However, it becomes
considerably wider in the case of disordered networks such as the random graphs or the
scale-free networks with the Poisson and the power-law degree distributions, respectively~\cite{
Albert2002}. In the previous paper~\cite{Kurzynski2014a}, we supposed, being primarily motivated
by the systems biology, that the network of states, like the protein interaction network and the
metabolic network, has evolved in the process of self-organized criticality~\cite{Sneppen2005}.
Such networks display a transition from the fractal organization on the small length-scale to
the small-world organization on the large-length scale~\cite{Rozenfeld2010}.

\begin{figure}[t]
\begin{center}
\includegraphics[scale=0.4,angle=-90]{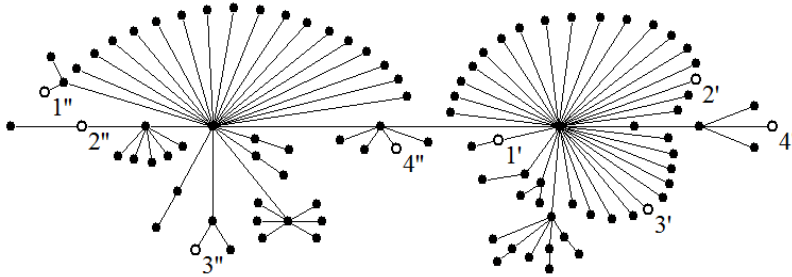}
\end{center}
\vspace{-0.5cm}
\caption{A sketch of the fractal scale-free network with the tree-like topology. Its size is
determined by the total number of $N\!=\!100$ nodes. The internal stochastic transitions and
the external jumps that break the detailed balance symmetry, assumed in Eq.~\ref{eq_7} for
$a\!\neq\!0$, are jointly described by the system of coupled master equations in Eq.~\ref{eq_2}.
The external transitions proceed through the gate composed of the pair of the selected nodes.
Their four exemplary configurations $(1',1'')$, $(2',2'')$, $(3',3'')$ and $(4',4'')$ are
shown on the network. Note that only a single gate is always to be used for each simulation
course.}
\label{fig_2}
\end{figure}

The exemplary network of states we have used in most of the cases in our studies is shown in
Fig.~\ref{fig_1}. As we demonstrate in Sec.~\ref{sec6}, its topology has a significant
impact on the statistical properties of the probability fluxes, apart from the general
properties predicted by the theoretical results in Sec.~\ref{sec5} and independent of the
network topology. Firstly, the characteristic tree-like topology is a result of the
well-thought-out construction originating from the theory of branching processes~\cite{Goh2006,
Kim2007,Harris1963}. The algorithm begins with a choice of a single node, corresponding to the
root of the growing tree. Then at each next step, each node from the previous generation,
also including the root in the first step, branches into $z$ descendant nodes with the
power-law probability function
\begin{equation}
\label{eq_11}
P(z)\varpropto z^{-\gamma},
\end{equation}
which is parameterized by the exponent $\gamma$. If the network displays the tree-like
topology~\cite{Kim2004} then no walk is possible, starting from and ending at the same
node without crossing each link belonging to the traced path at most once. Its second
noticeable feature is the presence of hubs, in this case, only the two highly connected
nodes. They have considerably higher degrees in comparison with the whole hierarchy of
the remaining nodes. Such an inhomogeneous pattern of the internode connections, but in
considerably larger networks, is the origin of the scale-free architecture, described
by the power-law degree distribution
\begin{equation}
\label{eq_12}
P(k)\varpropto k^{-\gamma}
\end{equation}
with the same exponent $\gamma$ as in Eq.~\ref{eq_11}. The network with $N\!=\!100$
nodes, shown in Fig.~\ref{fig_2}, is too small to prove this property, but the same
algorithm when applied to $N\!=\!10^{5}$ nodes generates the tree-like network, which besides
the scale-free topology also reveals other complementary properties like fractality~\cite{
Song2006,Rozenfeld2009}, self-similarity~\cite{Song2005} and branching criticality~\cite{Kim2007}.
The fractality results from the repulsion between the hubs, the self-similarity corresponds to the
invariance of the degree distribution function in Eq.~\ref{eq_12} under the network renormalization,
while the branching criticality denotes the presence of a plateau equal to unity in the mean
branching number dependence on the distance from the root.

Yet, it is well known that the scale-free networks that originate from another
algorithm, based on the growth and preferential attachment of the incoming nodes,
display the small-world pattern, because of the attraction between the hubs~\cite{
Barabasi1999,Watts1998}. At first sight, this property seems to be in evident
contradiction with the fractal topology. The small-worldness means that the average distance
between arbitrarily selected nodes depends only logarithmically on their total number,
whereas for fractal networks, it scales as the power-law with the number of nodes. This
ostensible discrepancy has been explained by the application of the renormalization-group
technique~\cite{Rozenfeld2010,Radicchi2008}. It appears that, upon linking the nodes of
the fractal scale-free tree via shortcuts with the distance $r$ distributed according to
the power-law function
\begin{equation}
\label{eq_13}
P(r)\varpropto r^{-\alpha},
\end{equation}
a transition to the small-world network occurs below some critical value of the exponent
$\alpha$. In the vicinity of its critical value, the network undergoes the topological phase
transition from the fractal to the small-world architecture. Close to the critical point it
is still the fractal, but only on the small-length scale, and manifests the small-world
characteristics on the large-length scale. We have used this method to dress the original
tree-like network in Fig.~\ref{fig_2} with the additional shortcuts to find the expected
differences between the steady-state distributions of probability fluxes resulting from a
global change in the network topology. The profound analysis of this effect and the other
reasons, such as the gate extension along the shortest path, the gate configuration on the
network and the very network size, having a direct impact on the statistical properties of
the stationary fluxes, are the objectives of Sec.~\ref{sec6}.

To supply the network with stochastic dynamics, described by Eq.~\ref{eq_2}, we assume the
probability of jumping from the currently occupied node $l$ to any of its $k_{l}$ nearest
neighbours to be identical in each time step and the mean time of jumping to be equal to
$\tau_{0}$. In consequence, the internal transition probability per unit time in Eq.~\ref{eq_6}
is inversely proportional to $k_{l}$:
\begin{equation}
\label{eq_14}
v_{l'l}=\frac{p}{\tau_{0}k_{l}}.
\end{equation}
As compared with Eq.~\ref{eq_8}, this formula contains an additional parameter $p$. To clarify
its meaning, let us combine at first Eq.~\ref{eq_14} with the detailed balance condition given by
Eq.~\ref{eq_7} and the normalization condition $\sum\limits_{l'}p^{\mathrm{eq}}_{l'}\!=\!1$. This
simple operation leads to the formula for the equilibrium occupation probability of the node $l$
\begin{equation}
\label{eq_15}
p^{\mathrm{eq}}_{l}=\frac{k_{l}}{\sum_{l'}k_{l'}}\!=\!\frac{k_{l}}{2L},
\end{equation}
where the summation runs over all the network nodes $N$, which gives the doubled number $L$
of the links. In the particular case of the connected tree $L\!=\!N\!-\!1$. A key issue in the
next step is an adequate choice of one of the two nodes, composing the gate from which the
external transition rate $v_{+}$ or $v_{-}$ (multiplied by $p$ as in Eq.~\ref{eq_14}) to the
second node, but without waiting on the first one, is supposed to be the fastest. In other words,
we select such a node $l\!=\!0'$ or $0''$ which for the predetermined transition time $\tau$
and the external force $a\!=\!0$ is occupied with the smallest equilibrium probability
$p^{\mathrm{eq}}_{l}$. Let us stress that for the input data to computer simulations we
always assume $a\geq 0$. Then, by virtue of Eqs.~\ref{eq_4},~\ref{eq_6},~\ref{eq_9},~\ref{eq_10}
and \ref{eq_14}, the final expression for the parameter
\begin{equation}
\label{eq_16}
p=\frac{1}{1+\tau_{0}(\tau p^{\mathrm{eq}}_{l})^{-1}}.
\end{equation}
As for the remaining nodes and the second node belonging to the gate, we have yet to complement
the transition (exit) probabilities from each of them to unity. This requires a determination
of the non-zero waiting probabilities $u_{ll}\!=\!\tau_{0}w_{ll}$ on these nodes. The parameter
$p$ was introduced to Eq.~\ref{eq_14} for a purely technical reason to only shorten the time of
computer simulations. Thus, it does not influence the numerical outcomes as well.

\section{Methodology of random walk simulations adapted to stationary fluxes}
\label{sec4}

We prove in Sec.~\ref{sec5} that despite the three general properties of the probability fluxes
emergent in any compact network (graph) equipped with a single gate, there is a need to determine
the two essential parameters that fully characterize their stationary distribution functions. They
can be defined within a specific theoretical model of a dynamics on a network (see Sec.~\ref{sec2})
or alternatively by numerical computations. In this paper, we support our further studies by means
of the Monte Carlo method, that is generally implemented in large-scale computer simulations~\cite{
Binder2010}.

A random walk process is a special variant of the Monte Carlo technique imitating
displacements of a hypothetical walker in a sequence of independent random steps,
triggered by the pseudo-random number generator~\cite{Klafter2011}. Importantly, this kind
of irregular wandering in a discrete space and time reflects all the essential features of the
Markovian stochastic dynamics we have assumed in our model. Moreover, if every node of
a network has a chance to be visited by the walker at least once, then its stochastic evolution
is said to be ergodic. By virtue of the ergodicity, the average over the evolution time of any
dynamical (random) variable is equal to the average over its stationary equilibrium distribution.
We apply this property in order to construct the steady-state distributions of the probability
fluxes. It is obvious that only an entirely compact network can make a sort of substrate for the
ergodic processes. 
\begin{figure}[t]
\begin{center}
\includegraphics[scale=0.45]{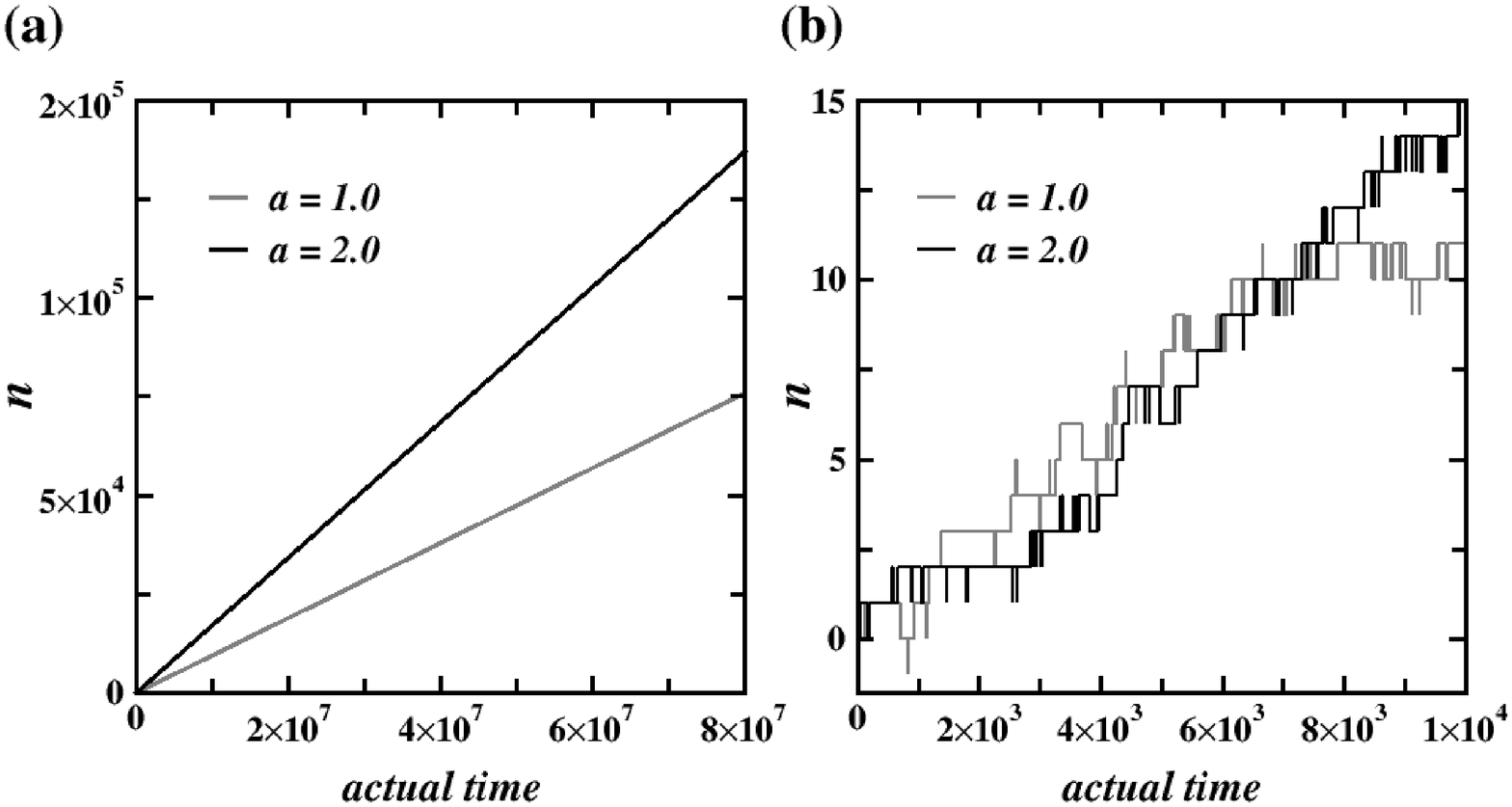}
\end{center}
\vspace{-0.5cm}
\caption{(a) The time course of the net number of external transitions on the fractal scale-free
network depicted in Fig.~\ref{fig_2} with the gate $(1',1'')$. The gray and the black data records
correspond to the actual time of $8\times10^7$ random walking steps. The slope of each line to the
horizontal axis determines the mean value of the probability flux. (b) The same results shown
in the shorter range of the actual time $10^4$ random walking steps. The repeating external
transitions through the gate are revealed as the characteristic straps corresponding to the shorter
or longer periods of time.}
\label{fig_3}
\end{figure}

When the walker moves around the network, it executes a series of random jumps between the neighbouring
nodes or makes external skips through the gate. All these displacements are dictated through the
internal, Eq.~\ref{eq_14}, and the external, Eqs.~\ref{eq_9} and \ref{eq_10}, transition rates.
Setting the unit of time $\tau_{0}\!=\!1$ in Eq.~\ref{eq_14} means that the actual time registered
during the simulation course is counted in the random walking steps. As a consequence, we neglect
in the output data the events related to the necessity of the waiting of the walker on the nodes. For
this reason, the actual time is completely independent of the parameter $p$ in Eq.~\ref{eq_14} and
the stochastic transitions between the network nodes are treated on an equal footing, except for
those passing through the gate. If the transition time $\tau$ through the gate is shorter than the mean
transition time $\tau_{l'l}$ between the neighbouring nodes, which, by virtue of Eqs.~\ref{eq_8},
\ref{eq_14} and \ref{eq_15}, is equal to the doubled number of links, i.e. $2L$, then the external
jumps are very frequent and the process is controlled by the internal dynamics. However, the more the
process is controlled by this dynamics, the higher the probability of waiting on the nodes. Accordingly,
the transitions through the gate must be overestimated, provided we exclude the waiting probabilities
from the data records. The opposite situation takes place when the external jumps are very rare in
comparison with the internal ones. Then, the process is controlled by the stochastic transitions through
the gate, which must be underestimated.
\begin{figure}[t]
\begin{center}
\includegraphics[scale=0.45]{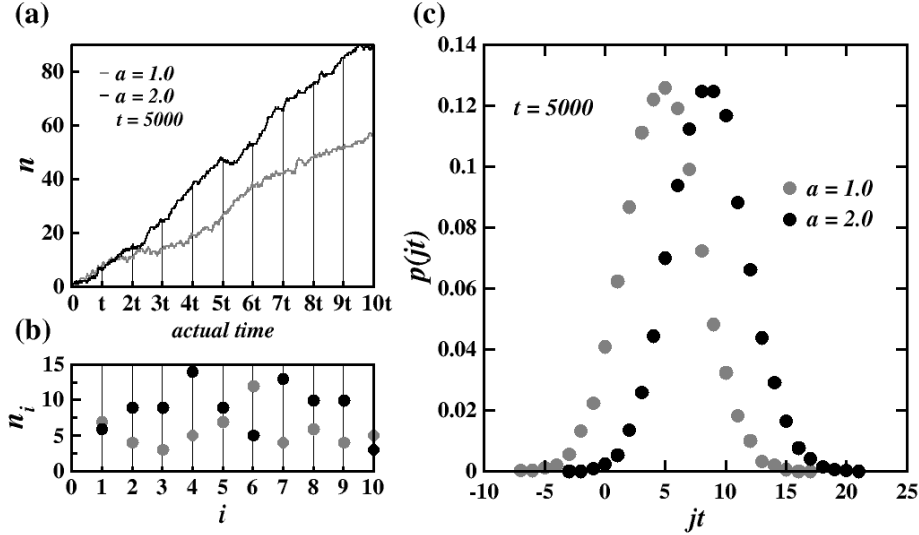}
\end{center}
\vspace{-0.5cm}
\caption{(a) The time course of the net number of external transitions on the fractal scale-free
network depicted in Fig.~\ref{fig_2} with the gate $(1',1'')$. The gray and the black data records
correspond to the actual time of $5\times10^4$ random walking steps and the two different constant
forces $a\!=\!1.0$ and 2.0, respectively. (b) The statistical sample of the first $h\!=\!10$ values
$n_{i}$ of the random variable $n$. (c) The histograms representing the stationary distributions
of the probability fluxes for the same time of averaging $t\!=\!5\times10^3$ random walking steps
and the aforementioned external forces obtained from the statistical sample of $1.6\times10^4$
values of the dimensionless random variable $jt$ (see text for a detailed description).}
\label{fig_4}
\end{figure}

Fig.~\ref{fig_3}a demonstrates two typical time courses of the net number $n$ of external transitions
through the gate $(1',1'')$ on the fractal scale-free tree depicted in Fig.~\ref{fig_2}. This
quantity is a sum of random steps that take two values $+1$ or $-1$ depending on the walker jumps
through the gate in a forward (from $1'$ to $1''$) or a backward direction, respectively. The
actual simulation time is equal to $8\times10^7$ random walking steps. The gray line was obtained
for the external force $a\!=\!1.0$, while the black one is the outcome for the larger force $a\!=\!2.0$.
The same external transition time $\tau\!=\!20$ was assumed in both simulation runs. Its value is lower
than the mean transition time $\tau_{ll'}$ between the adjacent nodes that, for the $N\!=\!100$ node tree
with $L\!=\!N\!-\!1\!=\!99$ links depicted in Fig.~\ref{fig_2}, equals $2L\!=\!198$ random walking steps.
This means that the passages through the gate are frequent and the process is controlled by the internal
dynamics of the system. We do not commit, though, too significant error in the determination of the
stationary distributions for the probability fluxes, counting the steps performed by the walker in the
actual time, because the time $\tau_{ll'}$ is the only one order of magnitude larger than the time $\tau$.
In the following, we will keep the value of the time $\tau$ unchanged. Fig.~\ref{fig_3}b displays the
same results as those shown on the adjacent plot, but in a much shorter range of the actual time,
amounting to $10^4$ random walking steps. This plot reveals the durations of the repeating external
transitions lasted till the walker leaves the gate. They are recognized as the characteristic straps of
different thicknesses, corresponding to the shorter or longer periods of the actual time.

The basic ideas of our methodology are collected in Fig.~\ref{fig_4}. The construction of
the steady-state distribution of the probability fluxes begins from a division of the total range
of the actual time into $h$ intervals of an optimally chosen length $t$ (see panel (a)). In this
example $h\!=\!10$ and $t\!=\!5\times10^3$ random walking steps. The term ''optimal'' is thought of
as referring to a determination of the sufficiently large statistical sample of the random numbers
$n_{i}$, where $1\leq i\leq h$. They are integers, obtained by adding up all the values $+1$ or $-1$
for the forward and the backward external transitions, respectively, within the $h$-th time interval
of the length $t$. The statistical sample for the first ten values $n_{i}$ of the random variable $n$,
or the probability flux $j$, taking the values $j_{i}\!\equiv\!n_{i}/t$, is shown on the panel (b).
Now, we can create a histogram counting how many times a particular value $j_{i}$ of the random flux
$j$ repeats itself in the statistical sample. The panel (c) contains two such histograms, constructed
for the time courses of the net number $n$ of the external transitions through the gate $(1',1'')$,
recorded in Fig.~\ref{fig_3}a. Note that the probability fluxes are the real numbers within the
range from 0 to 1 and their stationary distribution is determined with respect to the dimensionless
variable $jt$. Its first moment $J\!=\!\sum\limits_{i\!=\!1}^{h}j_{i}p(j_{i})$ defines the mean value
of the probability flux which corresponds to the slope of the straight lines fitted to numerical data
in Fig.~\ref{fig_3}a, when the actual time is long enough (by definition infinite).

\section{Verification of theoretical predictions}
\label{sec5}

\begin{figure}[t]
\begin{center}
\includegraphics[scale=0.35]{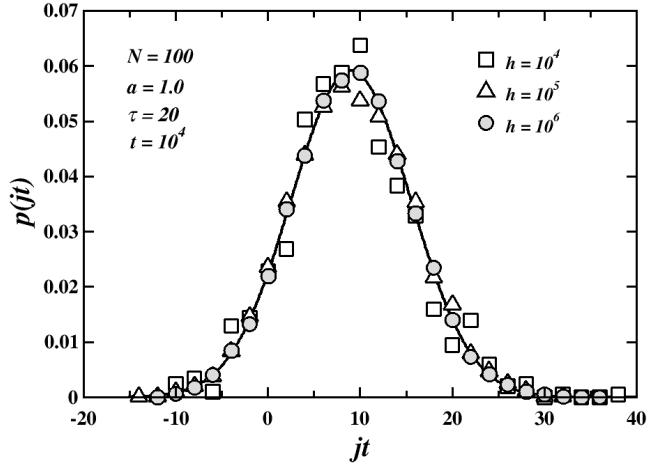}
\end{center}
\vspace{-0.5cm}
\caption{The quality test for the steady-state distributions of the probability
fluxes resulting from the stochastic dynamics on the network depicted in Fig.~\ref{fig_1}.
The external transitions through the gate $(1',1'')$ were exerted by the constant force
$a\!=\!1.0$. The optimal time of averaging $t\!=\!10^4$ random walking steps was assumed to
construct the statistical samples for $h\!=\!10^4$ (squares), $10^5$ (triangles) and $10^{6}$
(shaded circles) random variables $j$. The solid line representing the normal distribution
results from the fit to the shaded points that correspond to the largest statistical
sample.}
\label{fig_5}
\end{figure}

From the pragmatic point of view, the simulation time is always finite, thus a crucial question
arises about its duration having a decisive impact on the quality of numerical results.
Following our methodology, this question pertains actually to the size of the statistical
sample upon which the stationary distribution function for the probability fluxes is eventually
constructed. For this purpose, we have performed the random walk simulation in $10^{10}$ computer
steps on the network depicted in Fig.~\ref{fig_2} with the gate $(1',1")$ and the external constant
force $a\!=\!1.0$. Then, by appointing the time of averaging $t\!=\!10^4$ random walking steps, we
have divided the range of actual time into $h\!=\!10^{10}/10^{4}\!=\!10^{6}$ intervals in order to
construct three statistical samples of different sizes for the random variable $j$. Its steady-state
distribution functions are plotted in Fig.~\ref{fig_5} for $h\!=\!10^4$ (squares), $10^5$ (triangles)
and $10^{6}$ (shaded circles) time periods $t$. We see that the solid bell-like curve described
analytically by the normal (Gaussian) distribution
\begin{equation}
\label{eq_17}
p(j)=\frac{1}{\sqrt{2\pi}\Delta}\exp\biggl[-\frac{(j-J)^2}{2\Delta^2}\biggr]
\end{equation}
fits with a good accuracy only to the shaded circular points. In this formula the flux $j$ is
measured in the random walking steps $t$ of its determination, but in all the plots we define it as
the dimensionless variable. The above result is a direct consequence of a specificity of our model,
which we already mentioned in Sec.~\ref{sec2} and \ref{sec3}. Let us recall that the connectivity
of the network guarantees the (stationary) ergodicity of the process from which the equivalence
between the time and ensemble average arises. In turn, a network extended by a single gate ensures
the statistical independence (a lack of correlations) of the random variables $j$ whose probability
distribution converges through the central limit theorem to the Gaussian function~\cite{Jaynes2003}.

The function in Eq.~\ref{eq_17} is completely characterized by two moments, the average value of the
probability flux $J$ and the standard deviation $\Delta$. The average value estimated through the
fit of the Gaussian distribution to the numerical result is equal to $5.50\times10^{-4}$, while the
same value computed directly from the numerical data for $h\!=\!10^{6}$ equals $5.53\times10^{-4}$.
The error made in the determination of these values appears in second place after the decimal point.
Comparing the three results shown in Fig.~\ref{fig_5}, we can convincingly argue that the statistical
sample of $10^{6}$ values taken by the random variable $j$ is large enough to exploit it in
construction of the steady-state distributions of the probability fluxes with a good quality. In other
words, the choice of $t\!=\!10^{4}$ gives, thus, optimal time for $10^{10}$ simulation steps.
\begin{figure}[t]
\begin{center}
\includegraphics[scale=0.35]{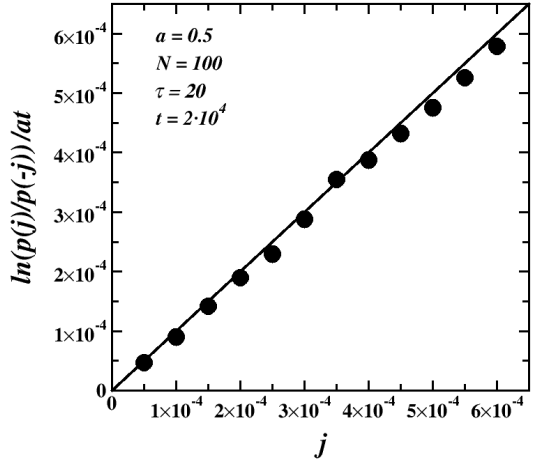}
\end{center}
\vspace{-0.5cm}
\caption{The stationary fluctuation theorem for the probability fluxes on the fractal tree-like
network in Fig.~\ref{fig_1}. No fit procedure has been performed between the computer experiment
(data points) and the theoretical result (the solid line).}
\label{fig_6}
\end{figure}

The non-zero width of the Gaussian distribution for the probability fluxes indicates that these
quantities are indeed the fluctuating variables. Their variations proceeding forward and backward
in time are related by the fluctuation theorem. If $p(j(t))$ denotes the stationary distribution
of the probability fluxes, depending on the time period $t$ of their determination, then the
stationary fluctuation theorem in the Andrieux-Gaspard form
\begin{equation}
\label{eq_18}
\frac{p(j(t))}{p(-j(t))}=\exp(jat)
\end{equation}
relates it to the distribution function $p(-j(t))$ of the opposite fluxes~\cite{Andrieux2006,
Andrieux2007c}. At equilibrium, when $a\!=\!0$, the external transitions through the gate in
the forward direction balance those in the backward direction, and hence the opposite fluctuations
of the probability fluxes are equiprobable. To verify that the fluctuation theorem in Eq.~\ref{eq_18}
applies to our model, we performed the test simulation of the random walk on the fractal
scale-free network with the gate $(1',1'')$ as before. Here, however, only the smaller external
force $a\!=\!0.5$ has been assumed. The stationary distribution of the probability fluxes was
determined for the time period of $t\!=\!2\times10^4$ random walking steps. The result shown in
Fig.~\ref{fig_6} reflects the linear form of the relation in Eq.~\ref{eq_18} upon representing it on
the semi-logarithmic plot. We emphasize that the solid line inclined at 45 degrees towards the
horizontal axis $j$ is not a consequence of the fitting procedure. More importantly, we also
performed additional test simulations and confirmed that the fluctuation theorem in Eq.~\ref{eq_18}
applies regardless of the network topology and the values of parameters assumed in our model.

The combination of Eqs.~\ref{eq_17} and \ref{eq_18} leads to a valuable conclusion, stating
that the standard deviation of the Gaussian distribution for the probability fluxes depends
on the square root of average flux $J$ for the predetermined values of the two parameters
$t$ and $a$:
\begin{equation}
\label{eq_19}
\Delta=\sqrt{\frac{2J}{at}}
\end{equation}
The central result of our paper enables us to relate the missing variable $J$ (the first moment in
Eq.~\ref{eq_17}) directly with the external constant force $a$ (see Eq.~\ref{eq_1}). In formal terms,
this relation is described by the nonlinear equation
\begin{equation}
\label{eq_20}
J=\frac{1-\mathrm{e}^{-a}}{J_{+}^{-1}+J_{-}^{-1}\:\mathrm{e}^{-a}},
\end{equation}
where the two crucial parameters $J_{+}$ and $J_{-}$ define the asymptotic values of the mean
flux $J$. These values can be determined in two ways, either within a certain theoretical model
of stochastic dynamics on a complex network with a single gate, or by computer simulations as is
shown in a moment. Thus, they completely characterize the two coefficients $J$ and $\Delta$ occurring
in the Gaussian distribution function for the probability fluxes $j$. As we have mentioned in the
introduction, Eq.~\ref{eq_20} was at first derived in the context of chemical processes for which
a total concentration of the reacting substrates is assumed to be conserved~\cite{Kurzynski2003,
Kurzynski2006}. In the present model, the conserved quantity is the probability flux.
\begin{figure}[tp]
\begin{center}
\includegraphics[scale=0.35]{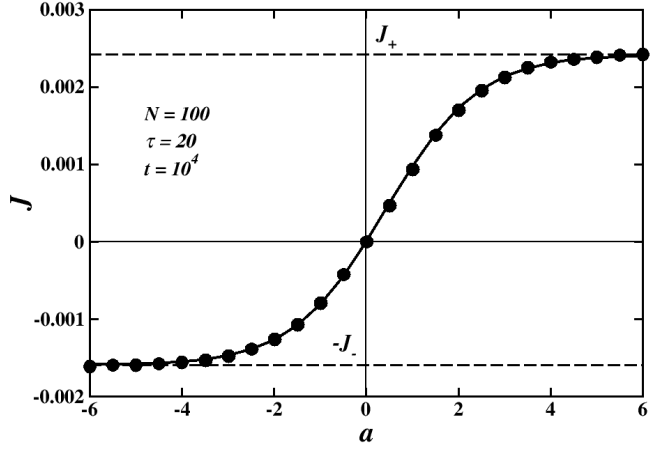}
\end{center}
\vspace{-0.5cm}
\caption{The dependence of the mean value of the probability flux $J$ on the external force $a$.
Two asymptotic values $J_{+}\!\approx\!0.0024$ and $J_{-}\!\approx\!-0.0016$ of the average
flux $J$ are distinguished by the dashed lines.}
\label{fig_7}
\end{figure}
\begin{figure}[bp]
\begin{center}
\includegraphics[scale=0.35]{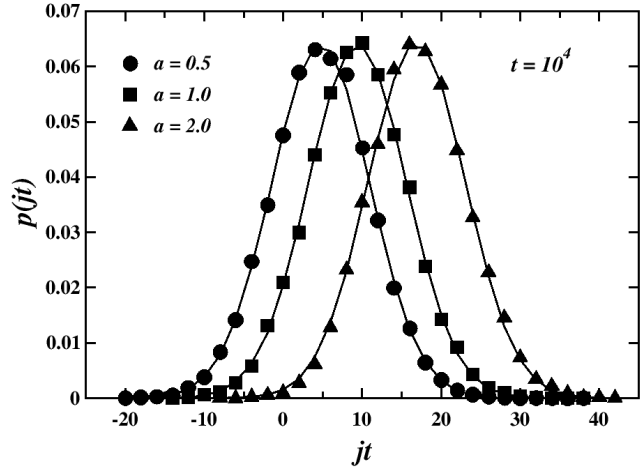}
\end{center}
\vspace{-0.5cm}
\caption{The displacement of the steady-state distribution function of the probability fluxes under
increasing external force $a\!=\!0.5$, 1.0 and 2.0. All simulations were conducted for the
stochastic dynamics on the tree-like network shown in Fig.~\ref{fig_2} with the fixed configuration
of the gate $(1',1'')$. The fluxes $j$ were averaged over the time period $t\!=\!10^4$ random
walking steps. The solid lines have the shape of a normal distribution and are the result of
the fitting procedure applied to numerical data.}
\label{fig_8}
\end{figure}

Fig.~\ref{fig_7} displays how the mean value $J$ of this flux depends on the external force $a$.
The points are the result of computer simulations we have conducted for the random walks on the
network depicted in Fig.~\ref{fig_2}. The external transitions through the single gate $(1',1'')$
were driven by force $a$ that was altered successively in the range from -6.0 to 6.0. The solid
curve is a fit of Eq.~\ref{eq_20} to the data points. Hence, we found the asymptotic values of the
two parameters $J_{+}\!\approx\!0.0024$ and $J_{-}\!\approx\!-0.0016$ marked in Fig.~\ref{fig_7}
by the dashed horizontal lines. Close to an equilibrium state, Eq.~\ref{eq_20} approximates to the
linear (Onsager) relation between the flux $J$ and the external force $a$:
\begin{equation}
\label{eq_21}
J\approx\frac{a}{J_{+}^{-1}+J_{-}^{-1}}.
\end{equation}
In this special case, $J(a)\!=\!-J(-a)$, but it does not remain arbitrarily true far from the
equilibrium state, unless $J_{+}\!=\!J_{-}$.

The steady-state distributions of the probability fluxes for the fixed time $t\!=\!10^{4}$ of their
determination and the three increasing values of the external force $a\!=\!0.5$, 1.0 and 2.0
are shown in Fig.~\ref{fig_8}. They correspond to the three points in Fig.~\ref{fig_5} to the
right of the vertical axis $a\!=\!0$. Their gradual displacement along the horizontal axis $j$ is a
consequence of the nonlinear dependence of the mean flux $J$ on the external force $a$ given by
Eq.~\ref{eq_20}. The continuous lines result from the fit of the normal distribution function to
the numerical data. We examined its standard deviation $\Delta\!\approx\!6.2\times10^{-4}$ and
found that this value is identical to five decimal places in all cases. Accordingly, $\Delta$
should not depend on the external force close to equilibrium state where $J\!\varpropto\!a$.

\begin{figure}[t]
\begin{center}
\includegraphics[scale=0.35]{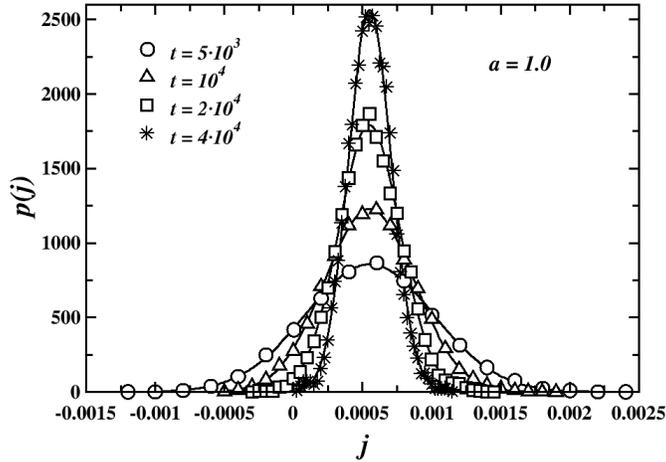}
\end{center}
\vspace{-0.5cm}
\caption{The steady-state distributions of the probability fluxes for the stochastic dynamics
on the tree-like network shown in Fig.~\ref{fig_2} with the gate $(1',1'')$ . Their widths are
determined by four different times of averaging $t\!=\!5\times10^3$, $10^4$, $2\times10^4$ and
$4\times10^4$ random walking steps according to Eq.~\ref{eq_20}. The external transitions
through the gate were powered by the same constant force $a\!=\!1.0$. The solid curves are
the normal distributions drawn (not fitted!) into numerical data on the basis of Eq.~\ref{eq_17}.}
\label{fig_9}
\end{figure}

Indeed, if we substitute the approximate formula in Eq.~\ref{eq_21} to Eq.~\ref{eq_19}, then the
standard deviation of the Gaussian function becomes inversely proportional only to the square root
of the time $t$:
\begin{equation}
\label{eq_22}
\Delta\approx\sqrt{\frac{2}{(J_{+}^{-1}+J_{-}^{-1})t}}.
\end{equation}
As Fig.~\ref{fig_9} demonstrates, the stationary distribution function of the probability fluxes
becomes more and more narrow in the vicinity of its growing maximum, when $t$ gradually increases.
In addition, its position remains unchanged with respect to the horizontal axis. It is particularly
important that all continuous lines in Fig.~\ref{fig_9} were traced (not fitted!) according to the
Gaussian function in Eq.~\ref{eq_17} with the two parameters being determined by Eqs.~\ref{eq_19} and
\ref{eq_20}. The results collected in this plot come from the random walk simulations performed on
the original tree-like network exposed in Fig.~\ref{fig_2}. The external transitions proceeding
through the gate $(1',1'')$ were exerted by the constant force $a\!=\!1.0$. However, the relation
$\Delta(t)\!\varpropto\!1/\sqrt{t}$ turns out to be fulfilled for any stochastic dynamics on a complex
network extended by a single gate because it results from the universal fluctuation theorem.

In the next section, we examine the additional effects that have an essential influence on the
stationary distributions of the probability fluxes. We exploit the results of the present
section to draw reasonable conclusions from this analysis.

\section{Factors that affect the average flux}
\label{sec6}

In the next three subsections we demonstrate how a network modification and a manipulation
of the gate affect the Gaussian distribution of the probability fluxes, altering its average
value $J$ and thereby the standard deviation $\Delta$ (see Eq.~\ref{eq_19}). These effects
are straightforwardly interpreted in terms of the previous theoretical results. However, the
nonlinear relation in Eq.~\ref{eq_20}, that connects the average flux with the external force,
also includes two independent parameters $J_{+}$ and $J_{-}$. A lack of a general method
prevents us from a determination of their values, so we support our further studies solely by
the computer simulations.

\subsection{Effect of shortcuts}
\label{subsec61}

We have noticed in Sec.~\ref{sec3} that the extension of the fractal scale-free tree with extra
linkages radically changes its architecture from the fractal topology on the small-length scale
to the small-world topology on the large-length scale~\cite{Rozenfeld2010}. To examine how this
transformation affects the mean value of probability flux $J$, we dressed the original network
in Fig.~\ref{fig_2}, having $N\!=\!100$ nodes and $L\!=\!N-1\!=\!99$ links, with additional
shortcuts. These connections were added to the network with probability $P(r)$ (see Eq.~\ref{eq_15})
depending on the shortest distance $r$ dividing each pair of the nodes. We assumed the value of
the exponent $\alpha\!=\!-2.5$. In this way, 211 of $(N(N-1)/2)-L\!=\!(N-1)(N-2)/2\!=\!4851$
all the possible shortcuts have been additionally attached to the tree-like network. Both the
configuration of the gate as well as the value of the external constant force $a\!=\!0.5$ remained
unchanged upon the network modification.
\begin{figure}[t]
\begin{center}
\includegraphics[scale=0.35]{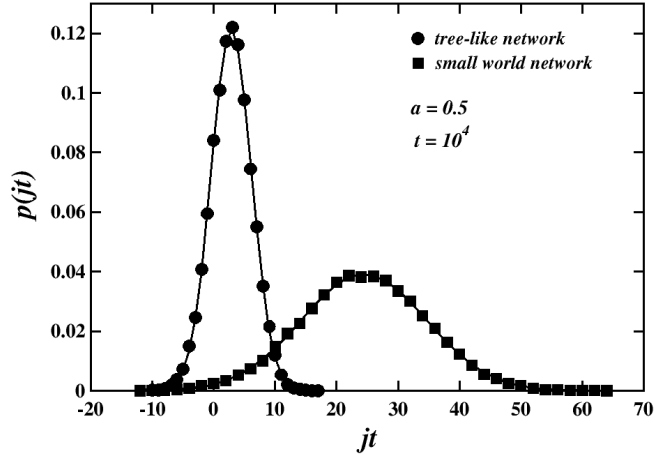}
\end{center}
\vspace{-0.5cm}
\caption{The influence of shortcuts on the stationary distribution of the probability
fluxes. The circles represent the numerical data for the random walk on the fractal
scale-free network with the tree-like topology, whereas the squares concerns results
for the same network, but dressed with additional shortcuts (see the text for a detailed
analysis). Both the solid lines were drawn into data points according to the Gaussian
distribution in Eq.~\ref{eq_17}.}
\label{fig_10}
\end{figure}

The steady-state distribution of the probability fluxes labeled in Fig.~\ref{fig_10} by the
dotted points is the outcome of the random walk simulation, performed on the fractal scale-free
tree. For comparison, its counterpart marked by the squared points originates from the random
walk simulation conducted on the small-world network. In both cases, the same time of averaging
$t\!=\!10^4$ random walking steps was adjusted. The shapes of continuous lines were defined on
the basis of the Gaussian function, Eq.~\ref{eq_17}, upon an earlier determination of the average
flux from Eq.~\ref{eq_20} and the standard deviation from Eq.~\ref{eq_19}, and then drawn into
the numerical data. The difference between the mean values of the probability flux results from
the fact that the long-range connections effectively shorten the distances between the nodes,
allowing the random walker to pass through the gate with much higher frequency. There are
alternative paths in the small-world network whereby the walker can visit the nodes more often
than on the tree-like network.
\begin{figure}[t]
\begin{center}
\includegraphics[scale=0.35]{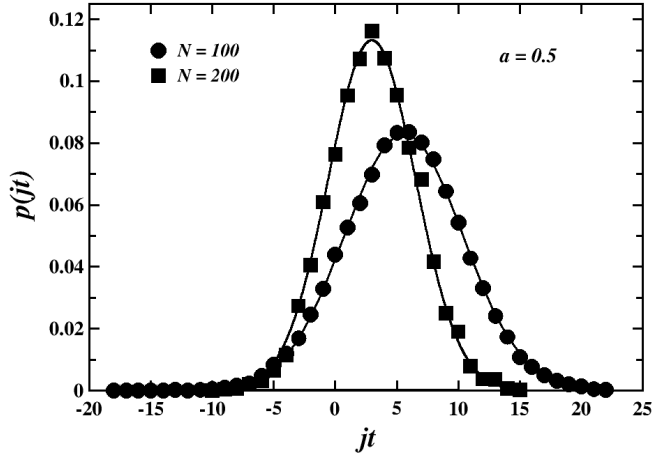}
\end{center}
\vspace{-0.5cm}
\caption{The steady-state distributions of the probability fluxes on the fractal scale-free
tree depicted in Fig.~\ref{fig_1} and its counterpart generated with $N\!=\!200$ nodes. In both
cases, the external transitions through the gate with equivalently distant nodes were powered by
the same constant force $a\!=\!0.5$. Also, the same time period of averaging $t\!=\!2\times10^4$
random walking steps was assumed. The solid lines are the Gaussian distributions in Eq.~\ref{eq_17}
drawn (not fitted!) into the data points.}
\label{fig_11}
\end{figure}

\subsection{Effect of the network size}
\label{subsec62}

A similar effect as the one involving the shortcuts is caused by the change in the network size.
To prove it, we have conducted the random walk simulations on two fractal scale-free trees of
different sizes. The first network with $N\!=\!100$ nodes was our original construction shown in
Fig.~\ref{fig_2}, while the second one, extended to $N\!=\!200$ nodes, was assembled according to
the same algorithm described in Sec.~\ref{sec3}. In both simulations, the external transitions
were proceeding through the gate composed of two equally distant nodes and were driven by the
identical constant force $a\!=\!0.5$. We have also assumed the same time of averaging
$t\!=\!2\times10^4$ random walking steps. The circles in Fig.~\ref{fig_11} correspond to the
numerical data for the steady-state distribution of the probability fluxes on the smaller
network, while the squares refer to that on the larger one. The solid curves have a profile
of the Gaussian functions and result from Eq.~\ref{eq_17} upon an earlier determination of the
two parameters, $J$ and $\Delta$ from Eqs.~\ref{eq_20} and~\ref{eq_19}, respectively.
\begin{figure}[p]
\begin{center}
\includegraphics[scale=0.35]{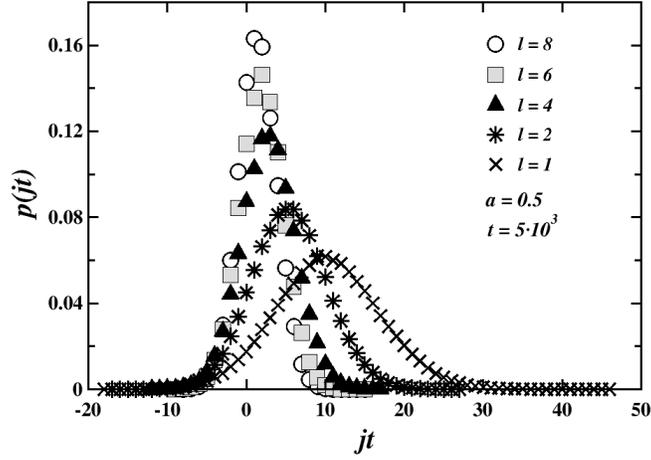}

\vspace{0.5cm}
\includegraphics[scale=0.35]{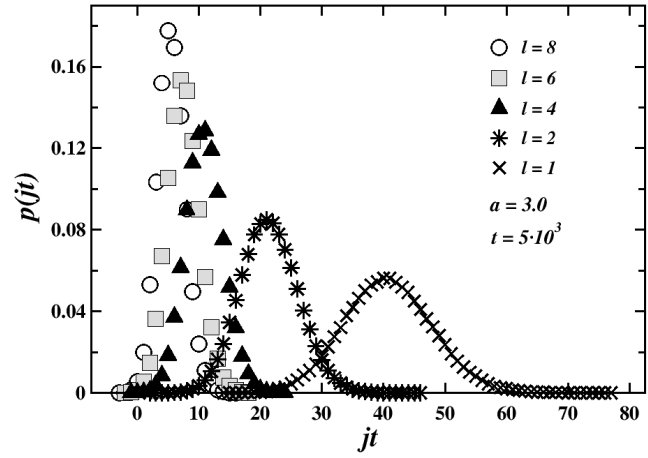}
\end{center}
\vspace{-0.5cm}
\caption{The impact of the gate extension on the steady-state distribution of the probability
fluxes. The distance $l$ between the nodes composing the gate was decreased due to their
displacements every one link from both ends of the network diameter. The external transitions
were driven by the same constant force $a\!=\!0.5$ (top) and 3.0 (bottom). Also, the same time
of averaging $t\!=\!5\times10^3$ random walking steps was assumed to construct the resulting
distribution functions.}
\label{fig_12}
\end{figure}

\subsection{Effects of the gate extension and configuration}
\label{subsec63}
\vspace{1mm}
The basic properties of the gate are a distance $l$ through links between its two constituent
nodes and their mutual positions relative to the other nodes composing a network. At first, we
deal with the former trait of the gate, referring to its second attribute in the later part
of this subsection.

In every connected network (graph) it is always possible to find at least one route that links
any two nodes. If a network has the tree-like architecture, such as the structure depicted in
Fig.~\ref{fig_2}, then each pair of the nodes is connected by a single path. Furthermore, there
always exists the longest path (not necessarily one), called a diameter, that separates the two
most distant nodes. The network depicted in Fig.~\ref{fig_2} contains only one diameter, extended
along the exposed horizontal line. It consists of eight links along which the two initially
outermost nodes composing the gate were gradually displaced in our tests. Their positions have
been changed before each simulation run in order to reduce a distance between them every one link
from both ends of the diameter. In this way, we performed in total ten independent random walk
simulations, establishing two different values $a\!=\!0.5$ and 3.0 for the external force. The
results for the steady-state distributions of the probability fluxes are collected in
Fig.~\ref{fig_12}. The same time of averaging $t\!=\!5\times10^3$ random walking steps was chosen
in both cases. When comparing the present results with those analyzed in the former sections, we
notice the essential similarities between them.

In this paragraph, we supplement our understanding of the second property of the gate concerning
its configuration on the network. The three exemplary deployments of the original gate $(1',1'')$
are shown in Fig.~\ref{fig_2}. For clarity, we distinguish them by the appropriate labeling
of the nodes that compose the gate (see the figure caption). Now, its size $l$ is fixed and
extended to five links for each configuration. Setting significantly different forces
$a\!=\!0.5$ and 3.0 and swapping the nodes between which the external transitions proceed, we
have performed a total of sixteen random walk simulations.  
\begin{figure}[p]
\begin{center}
\includegraphics[scale=0.35]{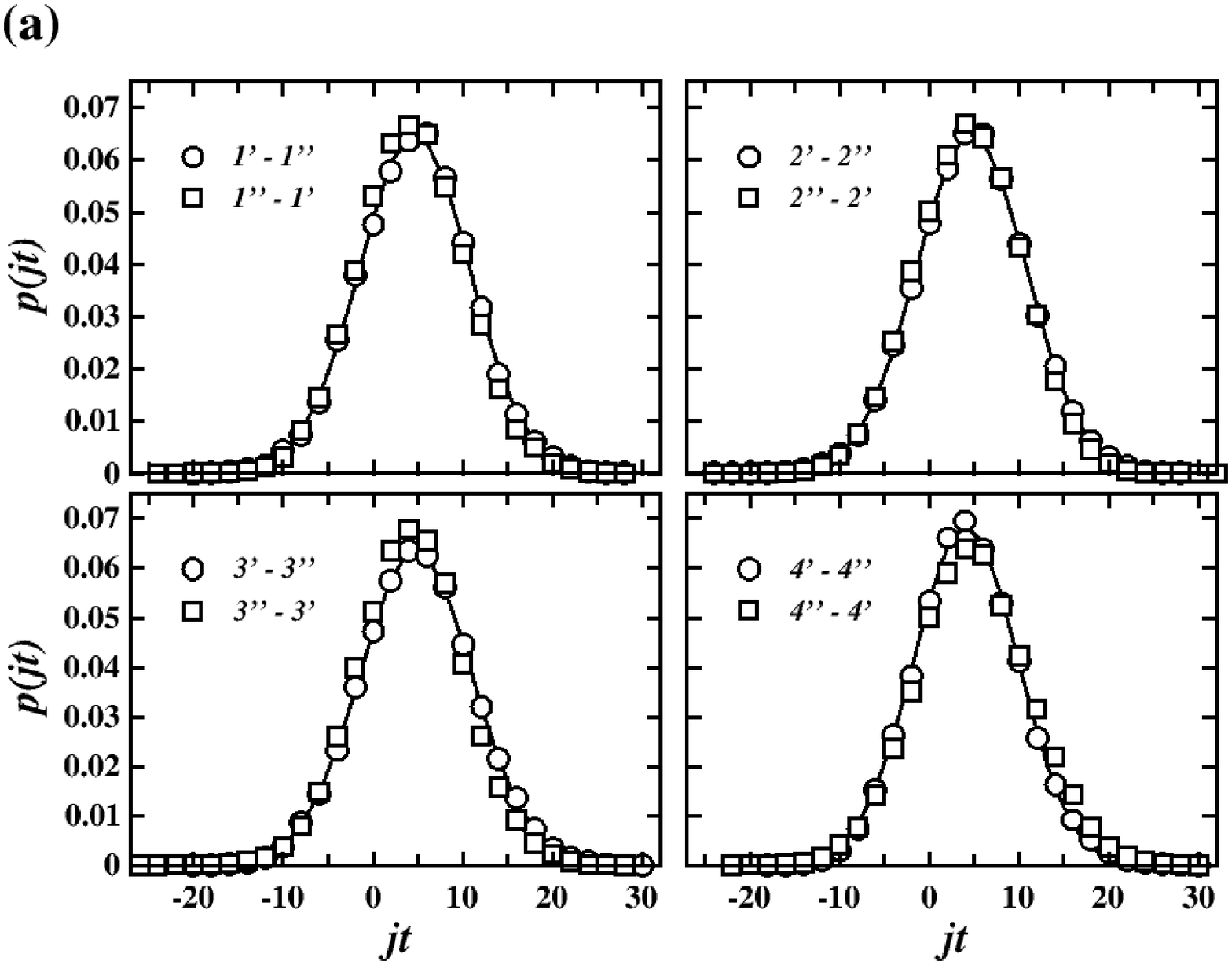}
\vspace{1cm}
\includegraphics[scale=0.35]{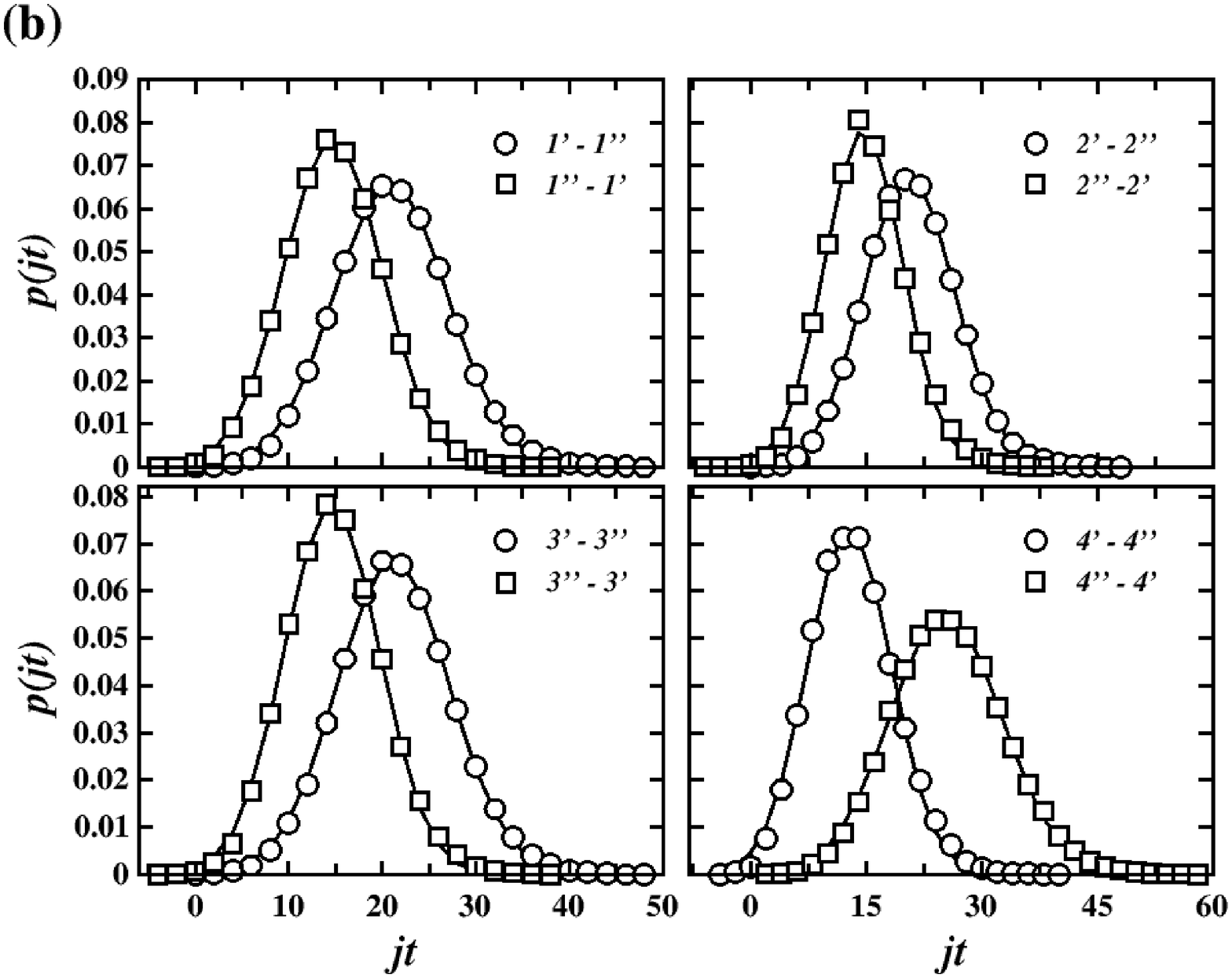}
\end{center}

\vspace{-1.5cm}
\caption{A comparison of the steady-state distributions of the probability fluxes emergent
from stochastic dynamics on the fractal scale-free tree for different configurations of the
gate (see Fig.~\ref{fig_2}). In all cases, the external transitions through the gate were
exerted by the same constant force $a\!=\!0.5$ (a) and 3.0 (b), and the time period of
averaging $t\!=\!10^4$ random walking steps was assumed. The continuous lines have the shape
of the Gaussian function given by Eq.~\ref{eq_17} and were drawn into the numerical results
after the determination of the two parameters defined in Eqs.~\ref{eq_18} and \ref{eq_20}.}
\label{fig_13}
\end{figure}
A permanent time of averaging $t\!=\!10^4$ random walking steps was supposed to construct
all the steady-state distributions of the probability fluxes. The results corresponding to
the smaller external force are collected in Fig.~\ref{fig_13}a. We find that the data points
contributing to the distribution functions converge on each other irrespectively of the
configuration of the gate and the replacement of the nodes. However, in the case of the
larger external force the data points assembled in Fig.~\ref{fig_13}b display the greater
divergence. The continuous lines obtained from the Gaussian distribution in Eq.~\ref{eq_17}
with two parameters determined separately by Eq.~\ref{eq_19} and~\ref{eq_20} were drawn and
do not fit into the data points.

To justify the differences between the results shown in Figs.~\ref{fig_13}a and \ref{fig_13}b,
we must refer to Eq.~\ref{eq_20} (see Fig.~\ref{fig_7}) from which the approximate relation
$J(a)\!=\!-J(-a)$ arises close to the equilibrium state. This is the reason why a swap of the
gate, or equivalently, an inversion of force $a$, cannot affect the average value $J$ of the
probability flux, regardless of its configurations on the network. But, far from the equilibrium,
when the force is large enough, the previous approximation is no longer satisfied what becomes
the cause of the discrepancies observed in Fig.~\ref{fig_13}b.

\section{Concluding remarks}
\label{sec10}

The keynote to our studies was the systematic analysis of the basic statistical
properties of the probability fluxes resulting, from the stochastic dynamics on
the fractal scale-free network extended by a single gate. We assumed that
this dynamics is described by the system of coupled master equations with the
transition rates inversely proportional to the node degree. Rather than solving
them analytically, we have supported our studies through the extensive computer
simulations. A method based on the random walk process allowed us to imitate the
displacements of the proverbial wonderer on the network and count its transitions
through the gate. These, in turn, were applied as the input data for calculations
of the probability fluxes and their stationary distributions.

The most important result of the present paper is the non-linear dependence of the mean
stationary flux on the external constant force that boils down to the Onsager linear
form close to the equilibrium state. The determination of the two parameters contained
in this relation allows us to specify both the average value and the standard deviation
of the stationary distributions of the probability fluxes. We examined that these distribution
functions converge to the normal (Gaussian) distribution irrespective of the size and topology
of the network, provided it is equipped with a single gate. Important in this regard is
the time of the determination of the stationary fluxes on which the standard deviation depends.
We demonstrated that a gradual elongation of this time transforms the Gaussian distribution
of the probability fluxes into the needle-like function centered around its average value.
The opposite effect, associated with the change in the mean value of the stationary
flux or equivalently with the displacement of the distribution function, is caused by the
variable external force, but its width remains constant in this case. Also, the other factors
affecting both the mean value and the width of the stationary distributions of the probability
fluxes, such as the addition of shortcuts to the tree-like network, the extension of the gate,
its configuration on the network and the network size, have been thoroughly studied and
interpreted in terms of the rigorous theoretical predictions.

The model of stochastic dynamics on the complex networks proposed in this paper is quite
general and presumably of the simplest form. Therefore, it may provide a starting point for
more advanced applications. For instance, by equating the network nodes with distinctive states
of some internally complex molecule we can consider its conformational dynamics, as opposed
to the vibrational one, in terms of the stochastic transitions between them. In the particular
case of the protein macromolecule, such a picture corresponds to the stochastic dynamics within
an ensemble of numerous conformational substates, composing its native state. The more extended
variant of this model, assuming two chemical reactions gated and controlled by the conformational
dynamics of the protein enzyme, is able to describe the process of the great biological importance,
that is the free energy transduction. Here, the first reaction (the input gate) donates the free
energy to the system due to the external thermodynamic force (the affinity) and the second one
(the output gate) consumes it, but partially because of dissipation, to perform work on some
external system. This problem was the subject of our previous papers, concerning the action of
the biological molecular machine~\cite{Kurzynski2014a,Kurzynski2014b}. There, however, we
considered the same type of stochastic dynamics as in the present paper. To make it more relevant
to conformation transitions within the protein enzyme, we should also assume that the transition
rates between the molecule states (the network nodes) depend on the temperature as well. We are
convinced that such a modification helps to raise the advantages of our model in the near future.

%\section*{Acknowledgments}
\bibliographystyle{elsarticle-num.bst}
\bibliography{netfluxes}

\end{document}